\documentclass[11pt]{article}
\usepackage{epsfig}
\usepackage{bm}
\usepackage{amssymb}
\usepackage{mathrsfs}
\usepackage{amsmath}
\usepackage{dsfont}

\begin{document}
\thispagestyle{empty}
$\;$ \hfill INT-PUB-12-030
\vspace*{7mm}

\begin{center}

{\LARGE 
Lattice study of the Silver Blaze phenomenon   \vskip3mm for a charged scalar $\phi^4$ field} 
\vskip12mm
Christof Gattringer$^{\, a,b}$,  
Thomas Kloiber$^{\, a}$
\vskip5mm
$^a\,$Karl-Franzens University Graz \\
Institute for Physics\\ 
Universit\"atsplatz 5, A-8010 Graz, Austria 
\vskip5mm
$^b\,$University of Washington, Seattle \\
Institute for Nuclear Theory \\
Box 351560, Seattle, WA 98195, USA
\end{center}
\vskip10mm

\begin{abstract}
We analyze a complex scalar field with $\phi^4$ interaction and a chemical potential $\mu$ on the lattice. 
An exact flux representation of the partition sum is used which avoids the complex 
action problem and based on a generalized worm algorithm we can run  Monte Carlo simulations at arbitrary densities. 
We study thermodynamical quantities as a function of the chemical potential $\mu$ for zero- and finite temperature. 
It is shown that at zero temperature thermodynamical observables  are independent of $\mu$ up to a critical value $\mu_{c}$
(Silver Blaze phenomenon). In a spectroscopy calculation we cross-check that $\mu_c$ agrees with the mass $m$ of the scalar field. 
The Silver Blaze region ends in a second order phase transition and we show that  for  
low temperatures the second order phase boundary persists and separates a pseudo Silver Blaze region from a condensed
phase with strong $\mu$-dependence.
\end{abstract}

\vskip5mm
\centerline{\sl Nuclear Physics B (in print)}

\vskip10mm
\noindent
{\tt christof.gattringer@uni-graz.at \\
thomas.kloiber@gmx.at }

\setcounter{page}0
\newpage
\noindent
\section{Introduction}
 
The QCD phase diagram is a topic which  currently  sees a lot of attention in experimental and theoretical work. A particularly interesting  (and rather unexplored) 
part is the region at low temperatures and medium to high densities (medium to high values of the chemical potential $\mu$). While for large chemical potential 
exotic states of matter are conjectured, for small chemical potential and vanishing temperature one expects physics to be independent of $\mu$: 
As long as the chemical potential is smaller than the mass of the state it couples to no excitations can be generated and physics must be independent of $\mu$. 
This is an interesting non-perturbative effect which has come to be known as Silver Blaze problem \cite{cohen}. 

In principle lattice QCD would be an adequate method to study such non-perturbative phenomena. However, when coupling a chemical potential 
the fermion determinant becomes complex and cannot be used as a probability weight in a Monte Carlo simulation. This so called complex action problem is a 
common feature in many field theories at non-zero $\mu$ and has considerably held up the study of QCD and other systems at finite density. Using new
ideas such as  worm algorithms \cite{worm}, designed for updating systems with local constraints, several interesting models at finite density were rewritten
in terms of a flux representation where the complex action problem is avoided and worm-based Monte Carlo simulations are possible (see, e.g., 
\cite{examples1,examples2,examples2b,examples2c,endres,thimble,examples3,wolff} for examples).  
This opens the possibility to explore the Silver Blaze phenomenon in an ab-initio lattice study.
 
In this article we analyze the Silver Blaze problem for a charged scalar $\phi^4$ field. For scalar fields a suitable flux representation 
\cite{endres,examples3,wolff} can be found 
which avoids the complex action problem at finite $\mu$ and Monte Carlo simulations at high densities become possible. The model 
has been analyzed previously using the complex Langevin approach and mean field techniques \cite{aarts1,aarts2}. In this paper we systematically study the transition at the
end of the Silver Blaze region at zero temperature and attempt to follow the transition line for non-zero temperatures. This work sheds light on Silver Blaze type of phenomena
and explores new strategies for overcoming the complex action problem in lattice calculations at finite density.

The article is organized as follows: In the next section we present the flux representation and our observables. In Section 3 the worm algorithm is described (see 
also the appendix for a pseudo-code listing) and tested. The numerical analysis at vanishing temperature is discussed in Section 4, followed by the results for 
non-zero temperature in Section 5. A summary and discussion section completes the paper. 
 
\section{Flux representation for the charged scalar field}

In the conventional representation the lattice action of the charged scalar field with a $\phi^4$-interaction is given by
\begin{equation}
S \; = \; \sum_x \!\left( \eta |\phi_x|^2 + \lambda |\phi_x|^4 - 
\sum_{\nu = 1}^4 \left[ e^{\mu \, \delta_{\nu,4} } \phi_x^\star \phi_{x+\widehat{\nu}}  \, + \,
e^{-\mu \, \delta_{\nu,4} } \phi_x^\star \phi_{x-\widehat{\nu}}  \right]  \right) .
\end{equation}
The first sum is over the sites $x$ of a $N_s^3 \times N_t$ lattice with periodic boundary conditions, the second sum is over the four 
Euclidean directions $\nu = 1,2,3,4$, and $\widehat{\nu}$ denotes the unit vector in $\nu$-direction. The degrees of freedom are the 
complex valued field variables $\phi_x$ attached to the
sites $x$ of the lattice.  $\eta$ denotes the combination $8 + m^2$, where $m$ is the bare mass parameter. The $\phi^4$ coupling 
is denoted by $\lambda$ and the chemical potential by $\mu$. All parameters are in units of the lattice spacing $a$, in other words the 
lattice spacing is set to $a = 1$ throughout this paper, except for some places where we display $a$ explicitly for clarity. 
The partition sum $Z$ is obtained by integrating the Boltzmann factor $e^{-S}$ over all field configurations, 
$Z = \int D[\phi] e^{-S}$. The measure is a product over all lattice sites $x$, with $\phi_x$ being integrated over the complex plane,
i.e., $D[\phi] = \prod_x \int_\mathds{C} d \phi_x / 2\pi$. The normalization with $2 \pi$ will be useful later. 

Note that in its present form the model has a complex action problem: For $\mu \neq 0$ the sum of the temporal nearest neighbor terms is complex. Thus 
the Boltzmann factor cannot be used as a probability weight in a Monte Carlo simulation. The problem can be overcome by using 
alternative degrees of freedom, i.e., a flux representation. The first step to obtain the flux representation of the partition sum is an 
expansion of the Boltzmann factors for the nearest neighbor terms 
\begin{eqnarray}
\hspace*{-6mm}&& \prod_{x,\nu} \exp\!\left( e^{\mu \, \delta_{\nu,4} } \phi_x^\star \phi_{x+\widehat{\nu}}\right) 
 \exp\!\left( 
e^{-\mu \, \delta_{\nu,4} } \phi_x \phi_{x+\widehat{\nu}}^\star \right)  =   
\\
\hspace*{-6mm}&& 
\sum_{\{ n, \overline{n}\}} \!\!
\left( \prod_{x,\nu}\! \frac{1}{n_{x,\nu}! \, \overline{n}_{x,\nu}!} \right) \!
\left(\! \prod_x e^{\,\mu [ n_{x,4} - \overline{n}_{x,4} ] } \right) \!
\left( \prod_{x,\nu}\! \Big(\phi_x^\star \phi_{x+\widehat{\nu}}\Big)^{n_{x,\nu}} \, 
\Big(\phi_x \phi_{x+\widehat{\nu}}^\star\Big)^{\overline{n}_{x,\nu}} \! \right) \!= 
\nonumber \\
\hspace*{-6mm}&& 
\sum_{\{ n, \overline{n}\}} \!\!
\left( \prod_{x,\nu}\! \frac{1}{n_{x,\nu}! \, \overline{n}_{x,\nu}!} \right) 
\left(\! \prod_x e^{\,\mu [ n_{x,4} - \overline{n}_{x,4} ] } \, {\phi_x^{\,\star}}^{\sum_\nu [ n_{x,\nu} + \overline{n}_{x-\widehat{\nu},\nu} ] } 
\, {\phi_x}^{\sum_\nu 
[ \overline{n}_{x,\nu} + n_{x-\widehat{\nu},\nu} ] }  \right) ,
\nonumber 
\end{eqnarray}
where by $\sum_{\{n,\overline{n}\}}$ we denote the sum over all configurations of the expansion variables $n_{x,\nu}, 
\overline{n}_{x,\nu} \in [0,\infty)$, i.e., the multiple sum
\begin{equation}
\sum_{\{n,\overline{n}\}} \; = \; \prod_{x,\nu} \, \sum_{n_{x,\nu} = 0}^\infty  \; \sum_{\overline{n}_{x,\nu} = 0}^\infty \; .
\end{equation}
The complex field variables are now written in polar form, $\phi_x = r_x\,e^{i \theta_x}$. Rewriting also the integration measure in 
polar coordinates one finds for the partition sum
\begin{eqnarray}
\hspace*{-3mm} Z &\!\!\! = \!\!\!& \sum_{\{ n, \overline{n}\}}  
\left( \prod_{x,\nu}\! \frac{1}{n_{x,\nu}! \, \overline{n}_{x,\nu}!} \right) \!\!
\left(\! \prod_x \int_{-\pi}^\pi \frac{d\theta_x}{2\pi} e^{-i\theta_x \sum_\nu [ n_{x,\nu}  - \overline{n}_{x,\nu} - 
( n_{x-\widehat{\nu},\nu}   - \overline{n}_{x-\widehat{\nu},\nu}) ] }\! \right)  \nonumber \\
&\!\!\!\times\!\!\!&\hspace{-2mm}
\left(\! \prod_x e^{\,\mu [ n_{x,4} - \overline{n}_{x,4} ] } \int_{0}^\infty\!\!\! dr_x \; r_x^{1 + \sum_\nu [ n_{x,\nu} + n_{x-\widehat{\nu},\nu} 
+ \overline{n}_{x,\nu} + \overline{n}_{x-\widehat{\nu},\nu} ] }  \; e^{-\eta r_x^2 - \lambda r_x^4} \right)\! . 
\end{eqnarray}
The integrals over the phase give rise to Kronecker deltas, which for notational convenience we will write as $\delta(n)$. The integrals over the modulus we abbreviate as
\begin{equation}
W(n) \; = \; \int_0^\infty dr \, r^{n+1} \,  e^{-\eta r^2 - \lambda r^4} \; .
\label{ww}
\end{equation}
They can easily be computed numerically. 
The partition sum now reads:
\begin{eqnarray}
Z &\!\! = \!\!& \sum_{\{ n, \overline{n}\}}  \left( \prod_{x,\nu}\! \frac{1}{n_{x,\nu}! \, \overline{n}_{x,\nu}!} \right) \!\!
\left(\! \prod_x \delta\left( \sum_\nu \big[ n_{x,\nu}  - \overline{n}_{x,\nu} - 
( n_{x-\widehat{\nu},\nu}   - \overline{n}_{x-\widehat{\nu},\nu}) \big] \! \right) \right)
\nonumber \\
& \!\! \times \!\!& \!\!
\left(\! \prod_x e^{\,\mu [ n_{x,4} - \overline{n}_{x,4} ] } \; W\!\left( \sum_\nu \big[ n_{x,\nu} 
+ \overline{n}_{x,\nu} + n_{x-\widehat{\nu},\nu} +  \overline{n}_{x-\widehat{\nu},\nu} \big]  \right) \right)\! . 
\end{eqnarray}
In this form the complex phase problem is gone. All weight factors for configurations of the $n$ and $\overline{n}$ 
variables are real and non-negative. However, configurations of the $n$ and $\overline{n}$ can still have weight 0. 
This is always the case if at a site $x$ the constraint enforced by the Kronecker deltas is violated, i.e., when at least 
for one $x$ one has  $\sum_\nu [ n_{x,\nu}  - \overline{n}_{x,\nu} - 
( n_{x-\widehat{\nu},\nu}   - \overline{n}_{x-\widehat{\nu},\nu}) ]  \neq 0$. 

In the current representation the constraints mix both, the 
$n$ and the $\overline{n}$ variables. The structure of the constraints can be simplified by introducing new variables
$k_{x,\nu} \in (-\infty,\infty)$ and $l_{x,\nu} \in [0,\infty)$. They are related to the old variables by
\begin{equation}
n_{x,\nu} - \overline{n}_{x,\nu} = k_{x,\nu} \qquad
\mbox{and} \qquad  n_{x,\nu} + \overline{n}_{x,\nu} = |k_{x,\nu}| + 2 l_{x,\nu} \; .
\end{equation} 
The partition sum turns into
\begin{eqnarray}
Z &\!\! = \!\!& \sum_{\{ k, l\}}  \left( \prod_{x,\nu}\! \frac{1}{(|k_{x,\nu}| + l_{x,\nu})! \, l_{x,\nu}!} \right) \!\!
\left(\! \prod_x \delta\left( \sum_\nu \big[ k_{x,\nu}  -  k_{x-\widehat{\nu},\nu}  \big] \! \right) \right)
\nonumber \\
& \!\! \times \!\!& \!\!
\left(\! \prod_x e^{\,\mu k_{x,4} } \; 
W\!\left( \sum_\nu \big[ |k_{x,\nu}| +  |k_{x-\widehat{\nu},\nu}| + 2( l_{x,\nu} + l_{x-\widehat{\nu},\nu}) \big]  \right) \right)\! . 
\label{Zfinal}
\end{eqnarray}
In the final form (\ref{Zfinal}), which we now refer to as flux representation, the constraints no longer mix the two types of flux variables. 
Obviously only the $k$-fluxes are subject to conservation of flux at each site $x$, i.e., only they must obey 
$\sum_\nu [ k_{x,\nu} - k_{x-\widehat{\nu},\nu}] = 0$ for all $x$.

Also observables need to be mapped to the flux representation. For thermodynamic observables which are obtained as derivatives of $\ln Z$ with respect to the 
various couplings this is particularly simple. We will need the following observables which we refer to as 
particle number density $n$, the particle number susceptibility $\chi_n$, the derivative $\chi_n^{\, \prime}$
of $\chi_n$ with respect to $\mu$, the field expectation value $\langle | \phi^2 | \rangle$, the corresponding susceptibility 
$\chi_{|\phi|^2}$ and its derivative $\chi_{|\phi|^2}^{\, \prime}$ 
with respect to  $\mu$. They are defined as
\begin{eqnarray}
\hspace*{-6mm}
&& 
n \, = \, \frac{T}{V}  \frac{\partial \, \ln Z}{\partial \mu} \,  = \,  \frac{1}{N_s^3 \, N_t } \frac{\partial\, \ln Z }{\partial \mu}  \; , \;
\chi_n \, = \,  \frac{\partial \, n}{\partial \mu} \; , \; 
\chi_n^{\, \prime} \, = \,  \frac{\partial \, \chi_n }{\partial \mu}   \; , 
\label{obslist} \\
\hspace*{-6mm}
&& 
\langle |\phi|^2 \rangle \, = \, \frac{-T}{V} \frac{\partial\, \ln Z}{\partial \eta} \, = \,  \frac{-1}{N_s^3 \, N_t } \frac{\partial \, \ln Z}{\partial \eta} \; , \;
\chi_{|\phi|^2} \, = \, \frac{-\partial   \langle |\phi|^2 \rangle  }{\partial \eta}   \; , \;
\chi_{|\phi|^2}^{\, \prime} \, = \, \frac{\partial  \, \chi_{|\phi|^2} }{\partial \mu} \, .
\nonumber
\end{eqnarray}
All observables are normalized by the spatial volume $V = N_s^3$ in order to make them intensive quantities. We remark, that the observables 
(\ref{obslist}) are dimensionful (except $\chi_{|\phi|^2 }$). Since we here use a fixed scale approach (the lattice spacing $a$ is set by $\eta$ and $\lambda$ and kept fixed)
and drive the temperature $T = 1/aN_t$ by changing $N_t$, we will often show results in lattice units. Where a comparison of results at different scales is necessary
we make quantities dimensionless using suitable powers of $\mu_{c}$, the critical value of the chemical potential at zero temperature (see the 
discussion below).

It is straightforward to work out the derivatives that define the observables (\ref{obslist}) 
also for the flux representation of $Z$ and we only discuss two examples to illustrate 
the general structure. Particularly simple is the particle number density. One immediately sees that the flux representation of $n$ is 
given by
\begin{equation}
n \; = \; \frac{1}{N_s^3\, N_t} \left\langle \sum_x k_{x,4} \right\rangle \; ,
\end{equation}
i.e., the particle number is simply the net flux of $k$ around compactified time (= the 4-direction). The expectation value $\langle ... 
\rangle$ is now of course understood as a vacuum expectation value in the flux representation. 
Not much more difficult is the flux representation for $\langle |\phi|^2 \rangle = - \partial/\partial \eta \ln Z / N_s^3 N_t$. Note that $\eta$ 
enters only through the weight factors $W(n)$ defined in (\ref{ww}) which have the obvious property 
$- \partial/\partial \eta W(n) = W(n+2)$. Application of the product rule to the product of the factors $W$ in (\ref{Zfinal}) and 
subsequent completion of the remaining products to the full Boltzmann factor gives rise to
\begin{equation}
\langle |\phi|^2 \rangle \; = \; \frac{1}{N_s^3\, N_t} \left\langle \sum_x \frac{W(f_x+2)}{W(f_x)} \right\rangle \, ,
\end{equation}
where we use the abbreviation $f_x = \sum_\nu  [ |k_{x,\nu}| +  |k_{x-\widehat{\nu},\nu}| + 2( l_{x,\nu} + l_{x-\widehat{\nu},\nu})  ]$.
In a similar way the other observables listed in (\ref{obslist}) can be obtained as higher moments of the fluxes and the sum of ratios
of $W$. 
 
\section{Worm algorithm and tests}
\label{sec_worm}

Having discussed the flux representation of the partition sum and the observables, we now address the task of finding a suitable 
algorithm. In the final representation (\ref{Zfinal}) of the partition sum we use two different sets of variables. One of them, the variables
$l_{x,\nu}$, are not subject to a constraint. For these we simply apply sweeps of local Metropolis updates: Each link $(x,\nu)$ is visited 
and we generate a trial configuration by increasing or decreasing $l_{x,\nu}$ by $\pm 1$ with equal probability. The change is accepted 
with the usual Metropolis probability given by the ratio of the weights of new and old configuration. A negative trial link is always 
rejected.

Considerably more interesting is the update of the flux variables $k_{x,\nu}$ where a suitable 
update needs to take into account the local 
constraints $\sum_\nu [ k_{x,\nu} - k_{x-\widehat{\nu},\nu}] = 0$. For that purpose we use a generalization of the Prokof'ev-Svistunov
worm algorithm \cite{worm}. The worm starts at a randomly chosen site of the lattice and then grows by adding links to form a chain of 
links. At each position $x$ the worm chooses randomly the next link from the 8 links attached to $x$. For that link it proposes to change 
the corresponding variable $k$ by $\pm 1$ and accepts that change with a local Metropolis step. 
At the starting point and the head of the worm the constraint is violated, and the worm must continue until its head reaches the starting 
point such that along the closed contour of changed links all constraints are intact again. A possible initial configuration for the $k_{x,
\nu}$ is to set $k_{x,\nu} = 0$ at all links, a configuration that clearly obeys the constraints.

The situation for a worm update of the variables $k_{x,\nu}$ is a little bit tricky: The local weight for a variable $k_{x,\nu}$ not only has 
factors that live on the link (the factorials in $1/(|k_{x,\nu}| + l_{x,\nu})!$ and the contribution from the chemical potential), but also two 
factors at the two endpoints of the link, the terms 
$W(f_x)$ and $W(f_{x+\widehat{\nu}})$ (where again $f_x = \sum_\nu  [ |k_{x,\nu}| +  |k_{x-\widehat{\nu},\nu}| + 2( l_{x,\nu} + l_{x-
\widehat{\nu},\nu})  ]$). The contributions from the factors $W$ at the endpoints need to be suitably distributed among subsequent steps 
of the worm. In the appendix we provide pseudo-code for the initialization of the worm and the subsequent steps, illustrating how we 
solved the weight assignment for the $W$ contributions. Worms for the variables $k_{x,\nu}$ are alternated with the local Metropolis 
sweeps for the unconstrained variables $l_{x,\nu}$.

In our simulations we use $N_s^3 \times N_t$ lattices with $N_s$ ranging from 4 to 24, and $N_t$ from 2 to 100. A full update sweep is 
the combination of one sweep for the $l_{x,\nu}$ variables and one worm for the $k_{x,\nu}$ fluxes. We typically use between 500.000 
and 1.000.000 measurements separated by 5 full update sweeps. The number of equilibration sweeps is typically 25.000. All errors we 
quote are statistical errors determined with the jackknife method. The weights $W$ defined in (\ref{ww}) were evaluated numerically
using Mathematica and pre-stored for the Monte Carlo simulation.

\begin{figure}[t!]
\centering
\hspace*{-2mm}
\includegraphics[width=1\textwidth,clip]{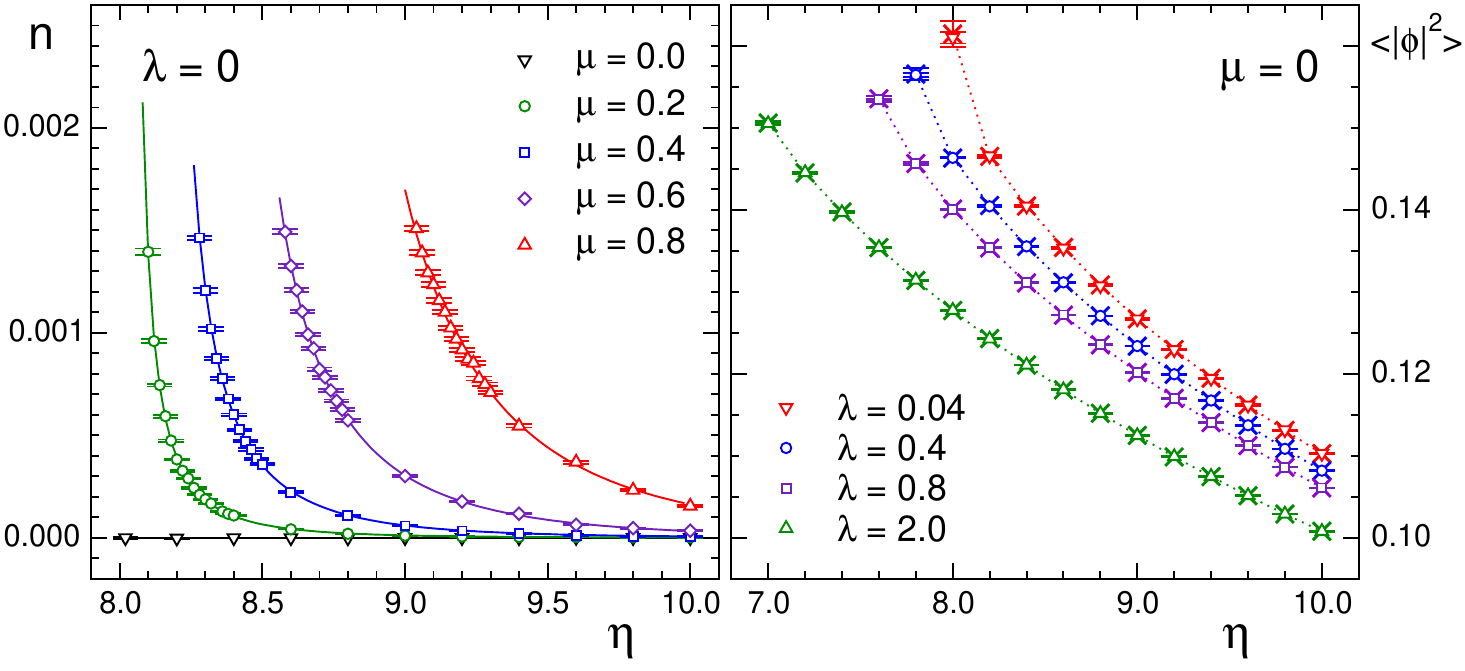}
\caption{Lhs.~plot: Comparison of results for the particle number density $n$ from the flux simulation at $\lambda = 0$ (symbols) to the 
exact result obtained from Fourier transformation (curves). We plot $n$  as a function of $\eta$ for different values of $\mu$ using 
lattices of size $8^4$. Rhs.~plot: Results for the expectation value $\langle |\phi|^2 \rangle$ as a function of $\eta$ for different values 
of $\lambda$, again on lattices of size $8^4$. The chemical potential is set to $\mu = 0$ and we compare the results from the flux 
simulation (triangles, circles, squares) to the outcome of a conventional $\mu = 0$ simulation (crosses).}
\label{checks}
\end{figure}

The algorithm was thoroughly tested by comparing it to the free case ($\lambda = 0$) where exact results can be obtained 
using Fourier transformation, and to a conventional simulation at $\mu = 0$. In Fig.~\ref{checks} 
we show examples for the comparison using $8^4$ lattices. The 
lhs.\ plot shows the particle number density $n$ as a function of $\eta$ for different values of the chemical potential $\mu$ for the 
case of $\lambda = 0$. The symbols are the data from our flux simulation and the full curves the results from Fourier transformation.  
The rhs.\ plot compares the  $\mu = 0$ results for the expectation value $\langle |\phi|^2 \rangle$ from the flux simulation (triangles) to 
those of a conventional simulation (crosses). We plot $\langle |\phi|^2 \rangle$ as a function of $\eta$ for different values of $\lambda
$. The two plots show that the results from the flux simulation very nicely match the exact curves for the free case and the data from a 
conventional simulation at $\mu = 0$. Further comparisons were done for other observables and more values of the parameters -- all 
with very good agreement. 

Finally we performed a comparison with the complex Langevin results at finite $\mu$ by Aarts
\cite{aarts1}, and we found excellent agreement of our data
with the complex Langevin results that are available on lattices of sizes $4^4$ to $10^4$.   
It is interesting to note that 
these results are very well described also by mean-field theory \cite{aarts2}. 
In that study the quartic coupling was fixed to $\lambda=1.0$, and the
boson mass $M$ satisfied $(a M)^2 = 1.47$. As a result of this heavy mass, the
temperatures reached on lattices $L^4 = 4^4$ to $10^4$ are in the range
$T/M = 0.08,.., 0.21$. At the same time, the system is large, since
$ML = 5,..,12$. So the observed mean-field behavior should come as no
surprise. Our comparison with \cite{aarts1,aarts2} confirms the findings by Aarts. 

To summarize: 
Our tests establish that the flux representation is correct, that the algorithm works, and that the 
observables were correctly mapped to the corresponding expressions in terms of the fluxes.

\section{Transition at zero temperature}

We begin our study of the phase diagram with an analysis of the situation at zero temperature. For this analysis we set the temporal extent of the lattice to 
$N_t = 100$, such that in lattice units the temperature is $T = 1/N_t = 0.01$. Although not exactly 0, this is much smaller than all other involved scales. To study 
the approach of our observables to the thermodynamic limit we analyze
different spatial volumes $N_s^3$ using $N_s = 4, 8, 12, 16, 20$ and 24. We consider two different combinations of the couplings $\eta$ and $\lambda$. 
The two points in parameter space are $\eta = 9.0, \, \lambda = 1.0$ and $\eta = 7.44, \, \lambda = 1.0$.  

\begin{figure}[p]
\centering
\hspace*{-6mm}
\includegraphics[width=1.1\textwidth,clip]{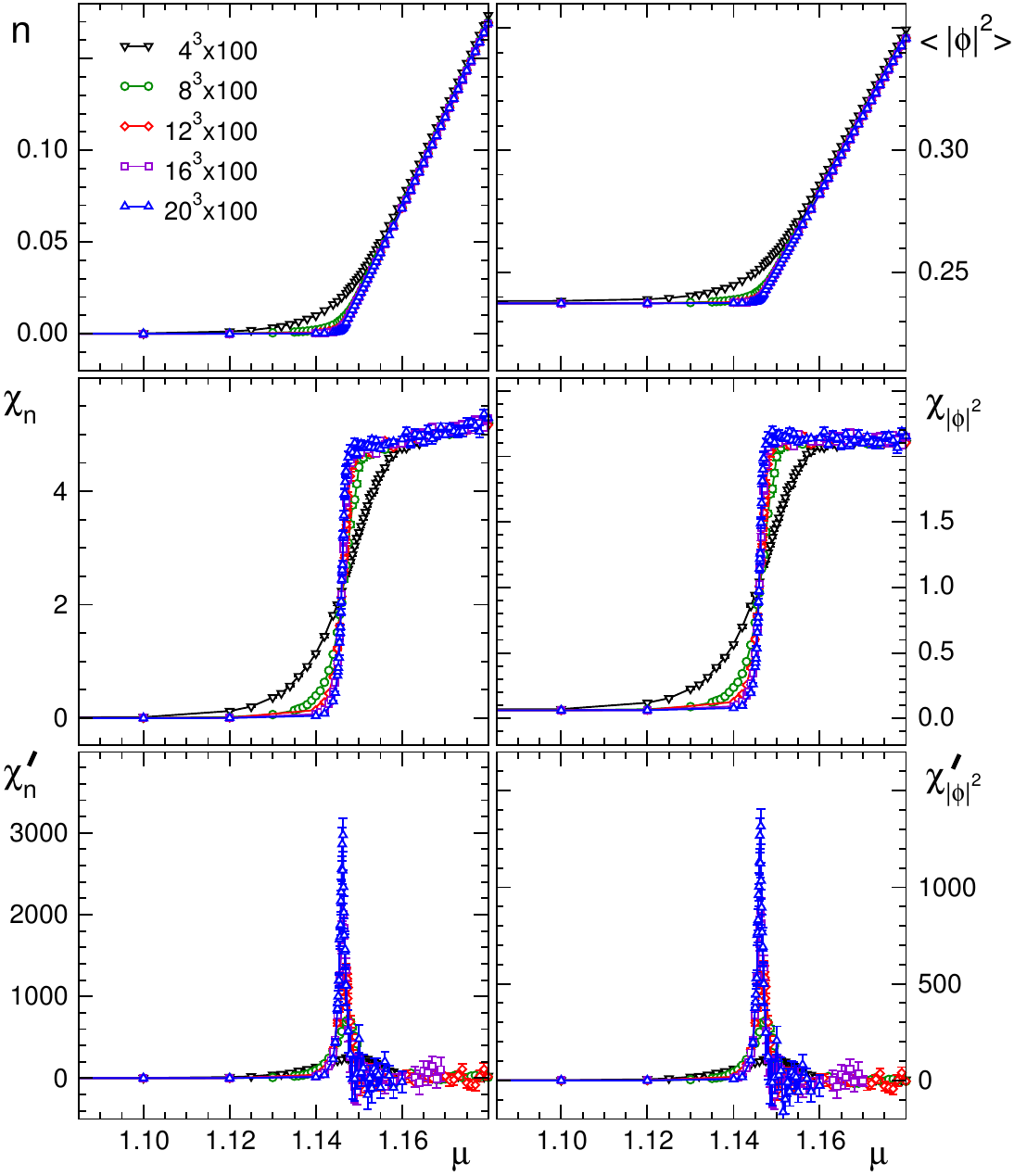}
\caption{Results at $T = 0.01$ for $\eta = 9.0$ and $\lambda = 1.0$. In the lhs.\ column of plots we show $n$, $\chi_n$ 
and $\chi_n^{\,\prime}$ as a function of $\mu$ (top to bottom). 
In the rhs.\ column we show $\langle |\phi|^2 \rangle$, $\chi_{|\phi|^2}$ and $\chi_{|\phi|^2}^{\, \prime}$. 
We compare the results for $4^3 \times 100, \, 8^3 \times 100,
\, 12^3 \times 100, \, 16^3 \times 100$ and  $20^3 \times 100$. }
\label{all_k450_T001}
\end{figure}

In Fig.~\ref{all_k450_T001} we show the  $\eta = 9.0$, $\lambda = 1.0$ results for $n$, $\chi_n$ and $\chi_n^{\,\prime}$ as a function of $\mu$  in the lhs.\ 
column of plots, and for  $\langle |\phi|^2 \rangle$, $\chi_{|\phi|^2}$ and $\chi_{|\phi|^2}^{\, \prime}$ on the rhs. The particle number density $n$ vanishes up to a 
critical 
value of $\mu_{c} = 1.146(1)$ which marks the end of the Silver Blaze region where observables are independent of $\mu$. 
For values $\mu > \mu_{c}$ the 
particle number density rises essentially linearly. Consequently the particle number susceptibility $\chi_n$ is discontinuous at $\mu_{c}$, jumping from  $\chi_n 
= 0$ to a finite value at $\mu_{c}$. The comparison of the different spatial volumes illustrates how the discontinuity emerges in the thermodynamic limit. 
Finally, when looking at the third $\mu$-derivative $\chi_n^{\, \prime}$ of the free energy we observe that for increasing volume a strongly 
peaked maximum emerges that very precisely marks the value of $\mu_{c} = 1.146(1)$. Furthermore, the height of the peak grows with increasing
volume $V = N_s^3$
(for a quantitative assessment see below), indicating  the emergence of a true singularity in the thermodynamic limit.
The observables $\langle |\phi|^2 \rangle$, $\chi_{|\phi|^2}$ and $\chi_{|\phi|^2}^{\, \prime}$ behave similarly to the derivatives with respect to $\mu$: The 
expectation value $\langle |\phi|^2 \rangle$ is constant below $\mu_{c}$ and for $\mu > \mu_{c}$ increases with constant slope. The corresponding 
susceptibility $\chi_{|\phi|^2}$ thus also develops a step in the thermodynamic limit. Finally, another derivative with respect to $\mu$, i.e., 
$\chi_{|\phi|^2}^{\, \prime}$, shows the corresponding peaks, which again grow with the volume.

\begin{figure}[p]
\centering
\hspace*{-6mm}
\includegraphics[width=1.11\textwidth,clip]{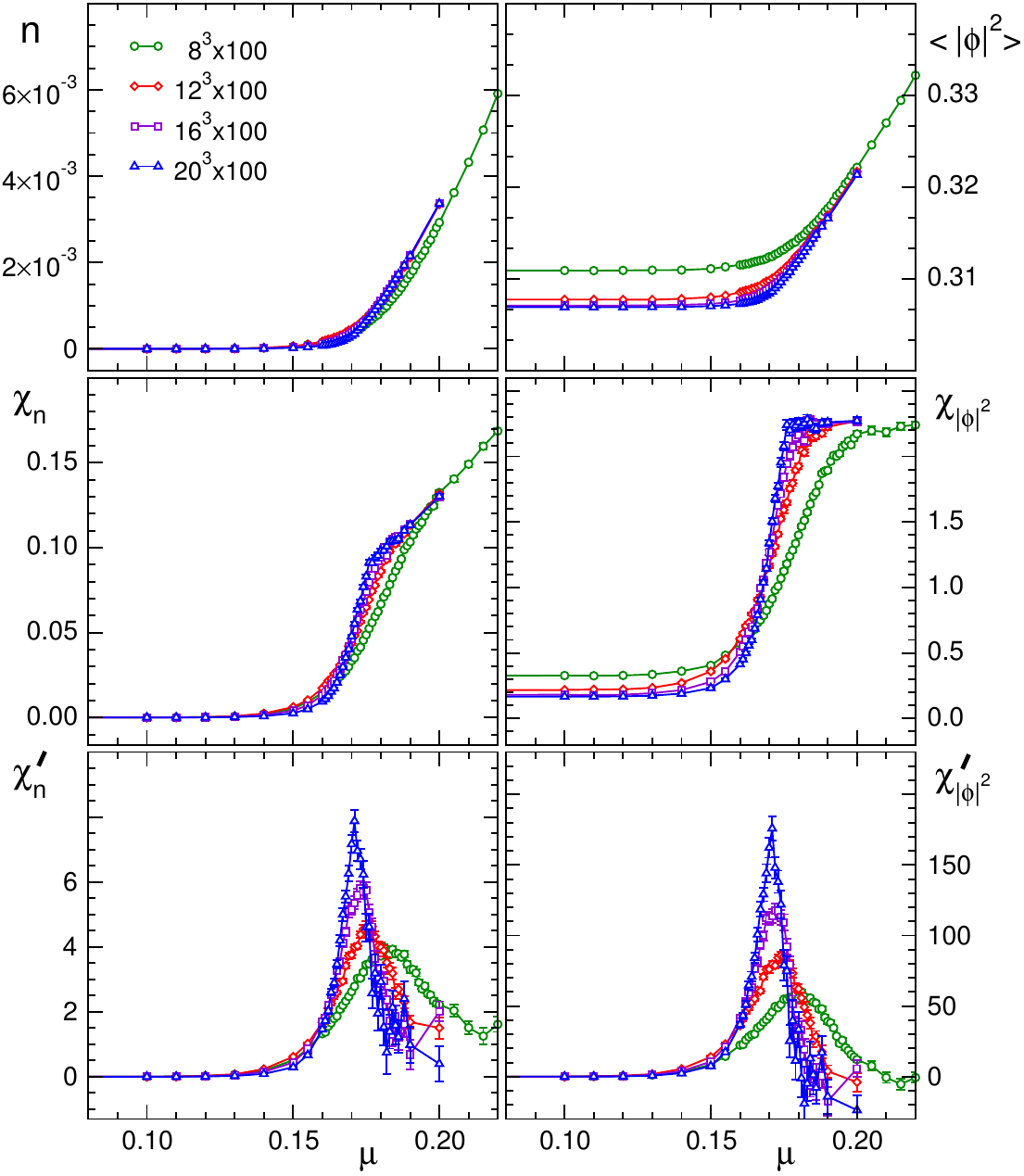}
\caption{Results at $T = 0.01$ for $\eta = 7.44$ and $\lambda = 1.0$. In the lhs.\ column of plots we show $n$, $\chi_n$ 
and $\chi_n^{\,\prime}$ as a function of $\mu$ (top to bottom). 
In the rhs.\ column we show $\langle |\phi|^2 \rangle$, $\chi_{|\phi|^2}$ and $\chi_{|\phi|^2}^{\, \prime}$. 
We compare the results for $8^3 \times 100,
\, 12^3 \times 100, \, 16^3 \times 100$ and  $20^3 \times 100$. }
\label{all_k372_T001}
\end{figure}

The results for the second parameter set, $\eta = 7.44$, $\lambda = 1.0$ are shown in Fig.~\ref{all_k372_T001}. Again we display $n$, $\chi_n$,
$\chi_n^{\,\prime}$, $\langle |\phi|^2 \rangle$, $\chi_{|\phi|^2}$ and $\chi_{|\phi|^2}^{\, \prime}$ as a function of $\mu$. In the plots we omit the 
data at $N_s = 4$ since 
for these the transition point in $\mu$ is shifted out of the $\mu$-range of our plots due to finite volume effects.  For the larger volumes the behavior of the 
observables is the same as for
the $\eta = 9.0$ runs, but now the critical value $\mu_{c}$ that marks the end of the Silver Blaze region is lower at $\mu_{c} = 0.170(1)$ in lattice units. Again we 
observe scaling of the maxima of the third derivatives with the volume and conclude that also at $\eta = 7.44$, $\lambda = 1.0$ we find a second order 
transition at  $\mu_{c} = 0.170(1)$. We remark that the drop of $\mu_{c} = 1.146(1)$ in lattice units at $\eta = 9.0$ to 
$\mu_{c} = 0.170(1)$ at $\eta = 7.44$ is  due to a considerable decrease of the lattice constant $a$. This implies that 
at $\eta = 7.44$ we work with a smaller lattice spacing $a$ and thus with a smaller physical spatial extent $a N_s$. Thus we expect to see more severe
finite volume 
effects at $\eta = 7.44$ than at $\eta = 9.0$. On the other hand with the finer lattice spacing we will be able to go to higher temperatures (see the next 
section).

To further confirm the finding of a singularity at $\mu_{c}$ we analyzed 
log-log plots of the height of the maxima of $\chi_n^{\, \prime}$ and $\chi_{|\phi|^2}^{\, \prime}$ versus the 3-volume $V = N_s^3$. 
For $\eta = 9.0$ the corresponding figure is Fig.~\ref{log_log_k450} with the $N_t = 100$ data we discuss in this section 
labeled by $T/\mu_c = 0.0087$, while the 
$\eta = 7.44$ results are shown in Fig.~\ref{log_log_k372} with the $N_t = 100$ data labeled by $T/\mu_c = 0.058$. 
The numerical results are represented by the symbols and we fit them with straight lines. The numbers next to the lines give their slope determined from the fits.
At $\eta = 9.0$ we find that the maxima of  $\chi_n^{\, \prime}$ and $\chi_{|\phi|^2}^{\, \prime}$ nicely follow a straight line 
with slopes that are close to 0.5, both for  $\chi_n^{\, \prime}$ and $\chi_{|\phi|^2}^{\, \prime}$. For $\eta = 7.44$ 
the finite volume effects are more severe, as discussed above, and we only show 
the data for our larger volumes, i.e, for $N_s = 12, 16, 20$ and $24$. Some of the data show an upward curving trend and if one uses $N_s = 16, 20$ and 24 for the fit,
one finds slopes of 0.39(4) for  the maxima of $\chi_n^{\, \prime}$ and 0.43(11) for $\chi_{|\phi|^2}^{\, \prime}$. 
Although for $\eta = 7.44$ the data are not as clean due to finite volume effects (and also the temperature in units of $\mu_{c}$ is not as low as for $\eta = 9.0$) the data indicate 
that also at $\eta = 7.44$ for the lowest temperature the maxima of $\chi_n^{\, \prime}$ and $\chi_{|\phi|^2}^{\, \prime}$ diverge as $V \rightarrow \infty$.

The analysis implies that at zero temperature the third derivatives diverge in the continuum limit and the susceptibilities develop a jump, indicating 
a second order transition at the end of the Silver Blaze region. 
In the first derivatives of the free energy 
the transition is characterized by the onset of a non-vanishing particle number density $n$ at $\mu_{c}$ and a linear response 
of the expectation value $\langle |\phi|^2 \rangle$ to changing $\mu$ for $\mu > \mu_{c}$. 

\begin{figure}[t]
\centering
\hspace*{-16mm}
\includegraphics[width=0.75\textwidth,clip]{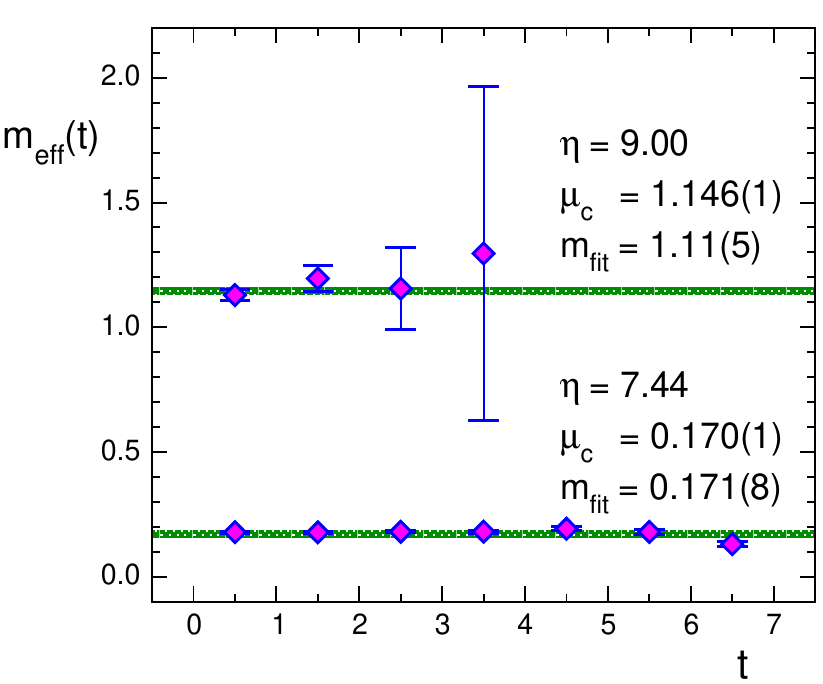}
\caption{Effective masses (symbols) for $\eta = 9.0$ (top set of data) and $\eta = 7.44$ (bottom). The effective masses are compared to the values of the
critical chemical potential $\mu_c$ (horizontal lines) and we annotate the values for $\mu_c$ as well as for the mass $m_{fit}$ obtained from an exponential fit of the 2-point function
(\ref{corr}). }
\label{effmass}
\end{figure}

We conclude the discussion of the $T = 0$ phenomenology with an important cross-check. The Silver Blaze regions ends at the point where the chemical potential becomes
equal to the mass $m$ of the excitation it couples to, in other words one expects $\mu_c = m$. In order to check this we performed a small spectroscopy calculation on 
$20^3 \times 40$ lattices using the conventional representation
at $\mu = 0$. The mass was extracted from the correlator of the zero momentum Fourier transform of $\phi$, i.e., the 2-point function
\begin{equation}
C(t) \; = \; \langle \, \widehat{\phi}(t) \, \widehat{\phi}(0)^* \,  \rangle \qquad \mbox {with} \qquad \widehat{\phi}(t) \; = \; \frac{1}{N_s^3} \sum_{\vec{x}} \phi(\vec{x},t) \; .
\label{corr}
\end{equation}
In Fig.~\ref{effmass} we show our results for the effective masses $m_{eff}(t+1/2) = \ln C(t)/C(t+1)$ of that correlator for both, $\eta = 9.0$ and $\eta = 7.44$. The effective masses are 
represented by the symbols and the horizontal lines are the respective values of $\mu_c$. It is obvious that the effective masses agree very well 
with the values of $\mu_c$\footnote{We remark that here the $\eta = 9.0$ data are more difficult to analyze, since the large mass $m = \mu_c = 1.146$ leads to a very rapid exponential 
decay of the correlator (\ref{corr}) and the relative 
statistical error quickly overlays the signal. Thus for $\eta = 9.0$ the effective mass plateau is rather short and the errors are large.}. An exponential fit of the correlator for $\eta = 7.44$ 
gives a mass of  $m_{fit} = 0.171(8)$ which is in excellent agreement with $\mu_c = 0.170(1)$. For $\eta = 9.0$ we find $m_{fit} = 1.11(5)$ which agrees with $\mu_c = 1.146(1)$ within error bars.
We conclude that we properly understand the Silver Blaze behavior at $T = 0$ .

\section{Non-zero temperature}

Having studied the transition at zero temperature we now explore how the transition line extends in the temperature direction of the phase diagram. As discussed, 
we work at fixed values of the 
couplings (again using our two values $\eta = 9.0$ and $\eta = 7.44$ at $\lambda = 1.0$), and drive the temperature by changing $N_t$. This has the 
advantage that we always work at a 
fixed scale and do not mix temperature and discretization effects, but on the other hand 
restricts the accessible temperatures to a discrete set. For our finite temperature studies we use the 
values $N_t = 100$, 20, 10, 9, 8, 7, 6, 5, 4, 3 and 2.
In lattice units the temperatures $T = 1/N_t$ thus range between $T = 0.01$ and $T = 0.5$. More interesting is the range of temperatures in units of the critical 
chemical potential at $T = 0$, i.e., 
the value $\mu_{c}$ which we determined in lattice units in the previous section\footnote{Whenever we use $\mu_c$ this refers to the
critical chemical potential at zero temperature.}. For $\eta = 9.0$,  where $\mu_{c}  = 1.146(1)$,
the largest temperature we can reach with $N_t = 2$ is 
$T/ \mu_{c} \sim  0.436$. For the finer 
$\eta = 7.44$ lattices ($\mu_{c} = 0.170(1)$) we can go considerably higher, 
up to $T/ \mu_{c} \sim  2.94$, although at the price of more severe finite volume effects. 

As the temperature is increased above zero, physics cannot be completely independent of $\mu$ below some  
critical value of the chemical potential. Still, at least for not too large temperatures, one 
should observe only very weak $\mu$-dependence of physical observables at sufficiently small chemical potential 
("pseudo Silver Blaze behavior"). In order to define the curves
in the phase diagram that separate the pseudo Silver Blaze behavior from a condensed phase characterized by 
strong $\mu$-dependence, we determined the position of the maxima of the third 
derivatives of the free energy
as a function of $\mu$ for different temperatures, using our $N_s = 20$ lattices for $\eta = 9.0$ and $N_s = 24$ for $\eta = 7.44$. 
For the $\eta = 9.0$ data we find that the positions of the maxima of $\chi_n^{\, \prime}$ and $\chi_{|\phi|^2}^{\, \prime}$ always agree 
exactly, while for the $\eta = 7.44$ data, where the finite size effects are more severe, we find agreement only within error bars. 

\begin{figure}[t]
\centering
\hspace*{-13mm}
\includegraphics[width=130mm,clip]{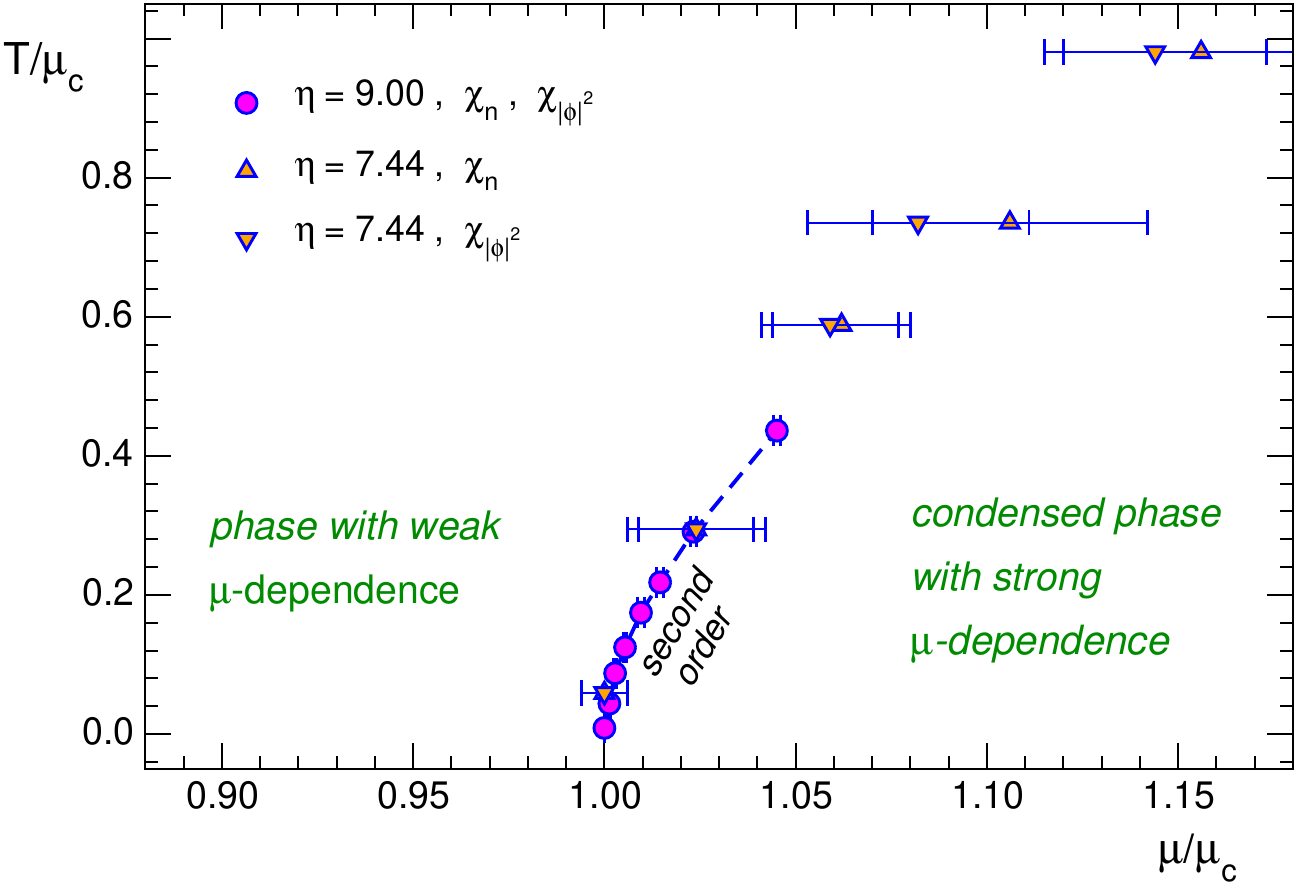}
\caption{Phase diagram determined from the positions of the maxima of $\chi_n^{\, \prime}$ and $\chi_{|\phi|^2}^{\, \prime}$ in the $\mu$-$T$ plane.
We combine data from $\eta = 9.0$ runs for $N_s = 20$ (circles connected with dashed lines) and $\eta = 7.44$ runs at $N_s = 24$ (triangles). 
$T$ and $\mu$ are expressed in units of $\mu_c$.}
\label{phasediagram}
\end{figure}

In Fig.~\ref{phasediagram} we show the positions of the maxima of $\chi_n^{\, \prime}$ and $\chi_{|\phi|^2}^{\, \prime}$
in the $\mu$-$T$ plane. For the $\eta = 9.0$ data we use filled circles connected with dashed lines (the plot also contains points from $\eta = 7.44$ 
(triangles) which we will discuss below).
We express both the temperature $T$ and the chemical potential $
\mu$ in units of the critical value $\mu_{c} = 1.146(1)$ determined in the zero temperature runs using $N_t = 100$. In these units the largest temperature we 
can reach with $\eta = 9.0$ is $T/\mu_{c} = 0.436$. When turning on the temperature, the phase boundary curves slightly to the right reaching 
$\mu / \mu_{c} = 1.045$ at $T/\mu_{c} = 0.436$, the highest temperature 
accessible for $\eta = 9.0$. This phase boundary separates pseudo Silver Blaze behavior from strong $\mu$-dependence.

\begin{figure}[t!]
\centering
\hspace*{-5mm}
\includegraphics[width=1.1\textwidth,clip]{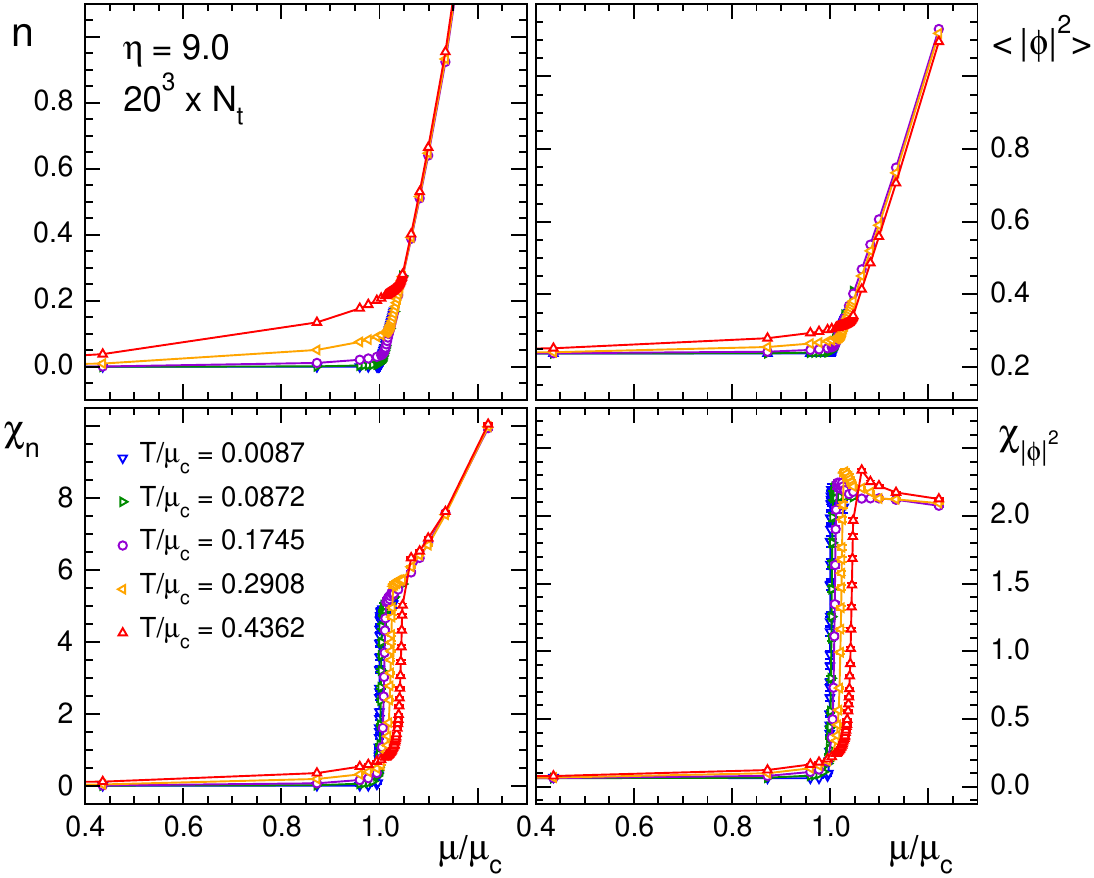}
\caption{Results for $n$, $\chi_n$, $\langle |\phi|^2 \rangle$ and $\chi_{|\phi|^2}$ for $\eta = 9.0$ and $N_s = 20$ as a function of $\mu$ for several 
temperatures corresponding to $N_t = 100, 
\, 10, \, 5, \, 3$ and 2. For $T$ and $\mu$ we use units of
$\mu_{c}$, while $n$, $\chi_n$, $\langle |\phi|^2 \rangle$ and $\chi_{|\phi|^2}$ are given in lattice units at a fixed scale 
(set by $\eta = 9.0, \lambda = 1.0$). }
\label{all_k450_allT}
\end{figure}

In order to assess the behavior across the phase boundary, in Fig.~\ref{all_k450_allT} we show the observables $n$, $\chi_n$, $\langle |\phi|^2 \rangle$ and 
$\chi_{|\phi|^2}$ from the $\eta = 9.0$ runs at $N_t = 20$ as a function of $\mu$ for several temperatures 
(again using for $T$ and $\mu$ units of $\mu_{c}$). In other words we 
show the observables along horizontal slices through the phase boundary in Fig.~\ref{phasediagram}.

While the first derivatives $n$ and $\langle |\phi|^2 \rangle$ were almost perfectly constant up to $\mu_{c} = 1.146(1)$ for our zero temperature runs 
(compare Fig.~\ref{all_k450_T001}), in Fig.~\ref{all_k450_allT} 
we now observe a deviation from constant behavior which becomes more prominent as the temperature is increased. However, the transition at the critical 
value of $\mu$ 
is still rather pronounced, and the corresponding susceptibilities $\chi_n$ and $\chi_{|\phi|^2}$ still seem to develop a discontinuity. 
The position of the discontinuity shifts towards larger values of $\mu$ in agreement with the curvature of the phase 
boundary in Fig.~\ref{phasediagram}.

\begin{figure}[t]
\centering
\includegraphics[width=110mm,clip]{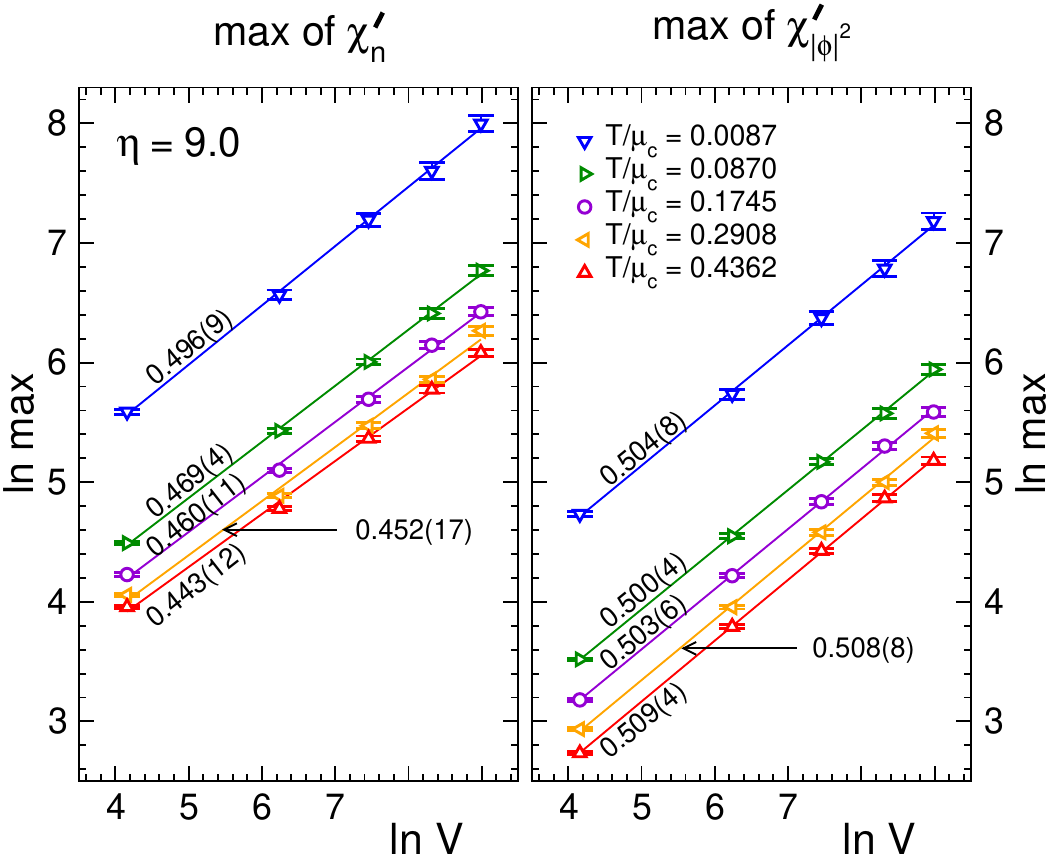}
\caption{Log-log plot of the maxima of the third derivatives of the free energy versus the 3-volume $V = N_s^3$ for $\eta = 9.0$ and $N_s = 4, 8, 12, 16$ and 20. 
We  compare different temperatures corresponding to $N_t = 100, \, 10, \, 5,\, 3$ and 2. 
The straight lines are linear fits to the data and the numbers near the lines are  the respective slopes.}
\label{log_log_k450}
\end{figure}

In order to analyze the order of the transition also at $T > 0$
we repeat the analysis of the volume dependence of the maxima of the third derivatives $\chi_n^{\, \prime}$ and 
$\chi_{|\phi|^2}^{\, \prime}$ as function of  the 3-volume $V$. The results of this analysis (done on $20^3 \times N_t$ lattices) 
for different temperatures are shown in 
Fig.~\ref{log_log_k450}. The symbols represent the numerical results and we fit them to straight lines, with the numbers next to the lines giving the corresponding 
slopes. It is obvious that the data are well described by a linear behavior with slopes near 0.45 for $\chi_n^{\, \prime}$ and slopes slightly above 0.5 for 
$\chi_{|\phi|^2}^{\, \prime}$. The plots show that also for our finite temperature values the third derivatives keep developing singularities 
in the thermodynamic limit. We conclude that for the temperature range we cover with our $\eta = 9.0$ runs, i.e., for temperatures up to $T/\mu_{c} = 
0.436$ a second order transition separates the pseudo Silver Blaze region from the condensed phase with strong $\mu$ dependence. 

We remark that the scaling behavior of the transition is expected to be different for the $T = 0$ and
$T > 0$ cases.  For $T = 0$ one expects that the transition is of mean field type,  which was indeed
confirmed by Aarts in \cite{aarts2} and our dual simulation perfectly reproduces this finding (as
already discussed in Section 3).  Due to dimensional reduction for $T > 0$ one expects a behavior in
the universality class of the 3-d XY model. However, it seems that it is difficult to explicitly 
extract this behavior from a lattice simulation with the resources available to us. This is reflected 
in the finding by Aarts \cite{aarts2} that  the
results on lattices ranging from $4^4$ to $10^4$ are equally well described by mean field behavior
which illustrates that the change in the critical behavior  must be very subtle. Also an
analysis in the continuum \cite{continuum}  shows that the non-linear contribution to $n$ versus $\mu
- \mu_c$ that is  predicted for finite $T$ is very small in the range of parameters we can access.
The finite volume analysis in Fig.~\ref{log_log_k450}  is not accurate enough to extract such a small
effect. 

An interesting question is of course where the phase boundary continues for larger temperatures. With the data taken at
$\eta = 9.0$ we reach a maximal  temperature of $T/\mu_{c} = 0.436$ and found the second order line to persist up to
that value. For temperatures above that number we need to use the  $\eta = 7.44$ configurations where $\mu_{c} =
0.170(1)$ in lattice units and temperatures of $T/\mu_{c} \sim 2.94$ can be reached. However, as  we have already
discussed, the $\eta = 7.44$ data suffer from more severe finite volume effects which spoil the accuracy of the
analysis (e.g.,  the peaks of the  third derivatives of the free energy become wider and it is more difficult to
exactly determine the phase boundary and the order of the transition).  Nevertheless we  try to  explore the trend for
the behavior along the phase boundary towards larger $T$. 

\begin{figure}[t!]
\centering
\hspace*{-4mm}
\includegraphics[width=1.09\textwidth,clip]{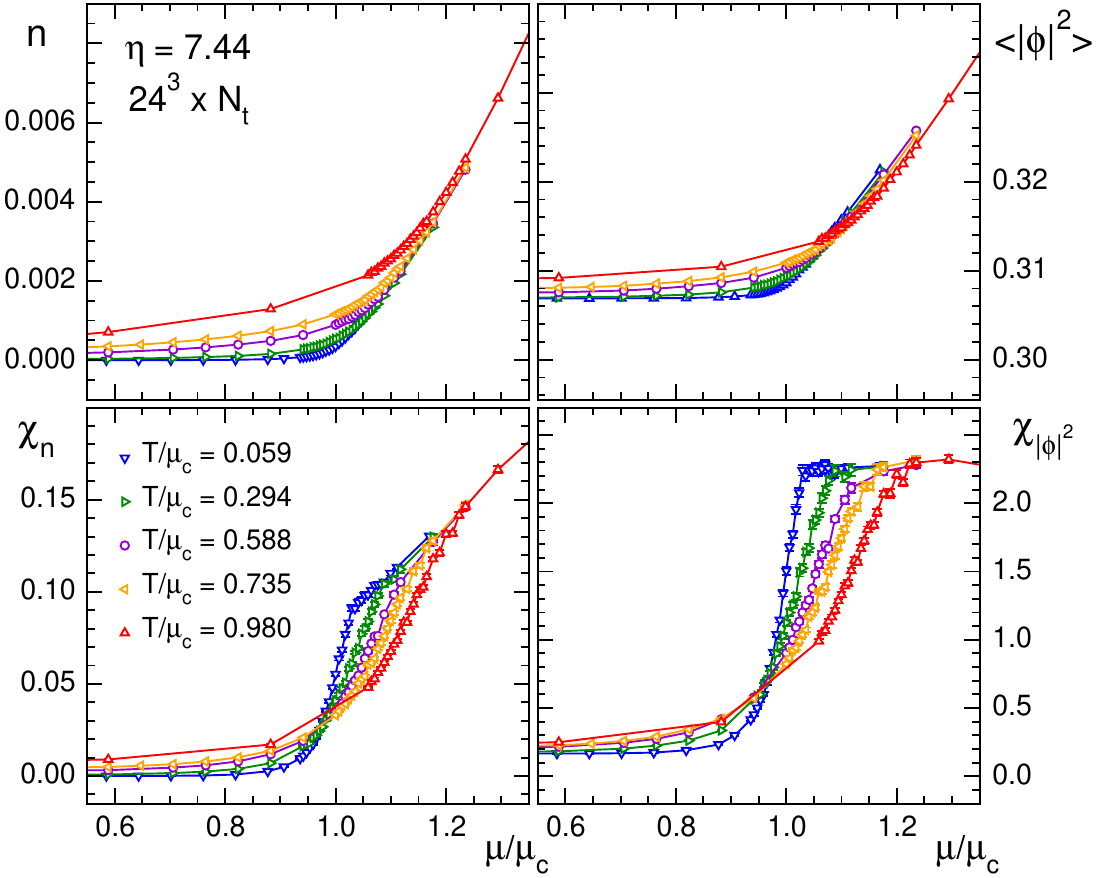}
\caption{Results for $n$, $\chi_n$, $\langle |\phi|^2 \rangle$ and $\chi_{|\phi|^2}$ for $\eta = 7.44$ and 
$N_s = 24$ as a function of $\mu$ for several temperatures corresponding to $N_t = 100, 
\, 20, \, 10, \, 8$ and 6. For $T$ and $\mu$ we use units of
$\mu_{c}$, while $n$, $\chi_n$, $\langle |\phi|^2 \rangle$ and $\chi_{|\phi|^2}$ are given in lattice units at a fixed scale 
(set by $\eta = 7.44, \lambda = 1.0$).}
\label{all_k372_allT}
\end{figure}

The phase diagram in Fig.~\ref{phasediagram} also contains a few points from $\eta = 7.44$ runs on $24^3 \times N_t$ lattices (triangles). As for $\eta = 
9.0$ we use the positions of the maxima of the third derivatives $\chi_n^{\, \prime}$ and $\chi_{|\phi|^2}^{\, \prime}$ to determine the phase boundary. 
However, while for $\eta = 9.0$ the positions of the maxima of $\chi_n^{\, \prime}$ and 
$\chi_{|\phi|^2}^{\, \prime}$ always match perfectly, for $\eta = 7.44$ this is the case only for temperatures up to $T/\mu_{c} \sim 0.5$. 
Consequently, at $\eta = 7.44$ we use different symbols for the maxima of $\chi_n^{\, \prime}$ (upward pointing triangles) and $\chi_{|\phi|^2}^{\, \prime}$
(triangles pointing downward). The two $\eta = 7.44$ points with $T/\mu_{c} < 0.5$ show matching maxima for $\chi_n^{\, \prime}$ and 
$\chi_{|\phi|^2}^{\, \prime}$ and fall on the curve set by the $\eta = 9.0$ data.  Those points confirm (although with larger error bars) the picture of a second order 
line up to at least 
$T/\mu_{c} = 0.436$ from the $\eta = 9.0$ runs. 
For the larger temperatures $T/\mu_{c} > 0.5$ the third derivatives  $\chi_n^{\, \prime}$ and 
$\chi_{|\phi|^2}^{\, \prime}$ peak at slightly different values. However, the positions of the maxima agree within error bars and we attribute the small discrepancy 
to the more severe finite volume effects for the $\eta = 7.44$ ensembles. The corresponding observables are shown in 
Fig.~\ref{all_k372_allT} where we plot $n$, $\chi_n$, $\langle |\phi|^2 \rangle$ and $\chi_{|\phi|^2}$ for $\eta = 7.44$ as a 
function of $\mu$ for several temperatures. 

\begin{figure}[t]
\centering
\includegraphics[width=110mm,clip]{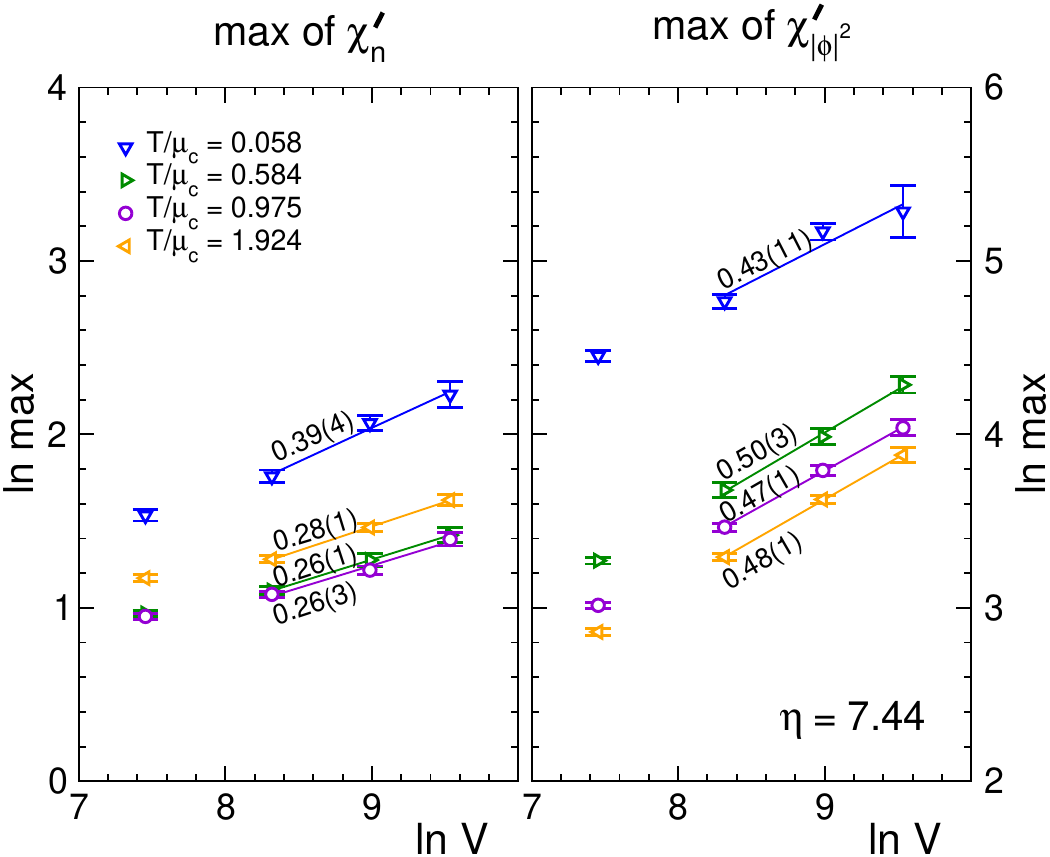}
\caption{Log-log plot of the maxima of the third derivatives of the free energy versus the 3-volume $V = N_s^3$ for $\eta = 7.44$ and $N_s = 12, 16, 20$ and 
24. We  compare different temperatures corresponding to $N_t = 100, \, 10, \, 6$ and 3. 
The straight lines are linear fits to the data (using only $N_s = 16, 20, 24$) and the numbers near the lines are  the respective slopes.}
\label{log_log_k372}
\end{figure}

In Fig.~\ref{log_log_k372} we repeat the finite volume study of Fig.~\ref{log_log_k450} now for $\eta = 7.44$.
We show the log-log plot of the maxima of $\chi_n^{\, \prime}$ and $\chi_{|\phi|^2}^{\, \prime}$ versus the spatial volume $V = N_s^3$. It is obvious that here the 
data do not as nicely fall on a straight line as in the corresponding plots for $\eta = 9.0$ shown in Fig.~\ref{log_log_k450}. We fit only the data points for the three 
largest volumes and as in Fig.~\ref{log_log_k450} denote the slopes next to the lines. It is quite interesting to observe that the slope for the maxima of 
$\chi_{|\phi|^2}^{\, \prime}$ remains at values close to 0.5 up to the largest temperatures, indicating that the susceptibility $\chi_{|\phi|^2}$ manages to 
develop a discontinuity in the thermodynamic limit also for our highest temperatures. The stronger finite volume effects at $\eta = 7.44$ 
make the analysis more difficult for $\chi_n^{\, \prime}$: The slopes for  $\chi_n^{\, \prime}$ are smaller for the smaller volumes 
but show a tendency towards pointing upwards for larger $V$, indicating second order behavior for the $\mu$-derivatives also up
to the highest temperatures we can access. 

\begin{figure}[t!]
\centering
\hspace*{-10mm}
\includegraphics[width=0.75\textwidth,clip]{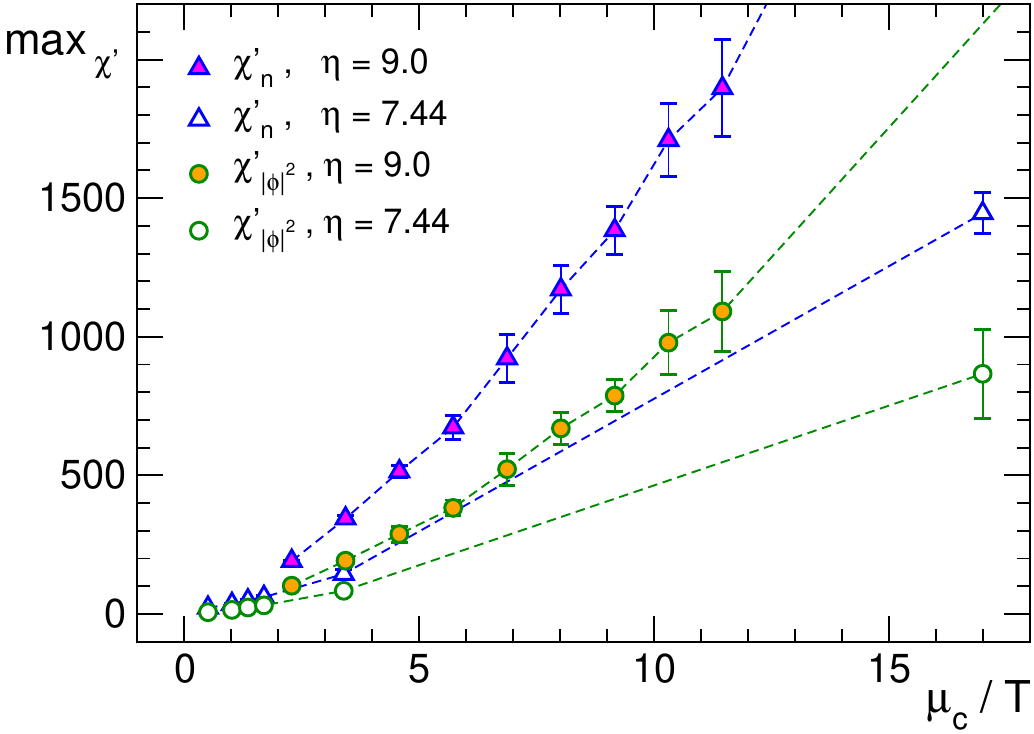}
\caption{Size of the maxima of the third derivatives $\chi_n^\prime$ and $\chi_{|\phi|^2}^\prime$ (both made dimensionless
with suitable powers of $\mu_c$) as a function of the inverse temperature in units of $\mu_c$. We compare the 
results for $\eta = 9.0$ on $20^3 \times N_t$ lattices and $\eta = 7.44$ on $24^3 \times N_t$. }
\label{step}
\end{figure}

One expects that the second order line persists for large $T$ due to the breaking of the U(1) symmetry at the
transition to the condensed phase.  In Fig.~\ref{step} we attempt a plausibility test for this scenario: We plot the
height of the maxima of the third derivatives $\chi_n^\prime$ and $\chi_{|\phi|^2}^\prime$ (both made dimensionless
with suitable powers of $\mu_c$) as a function  of the inverse
temperature in units of $\mu_c$. We find that the maxima are non-zero at finite $T$ and only for $1/T
\longrightarrow 0$ the curves reach 0. Furthermore the maxima for the 
$\eta = 9.0$ runs are larger than those at $\eta = 7.44$ as expected from the larger volume at $\eta =
9.0$. Although this analysis is only an exploratory test at a fixed volume, it supports
the existence of a phase transition throughout the phase diagram.    

\begin{figure}[t!]
\centering
\hspace*{-5mm}
\includegraphics[width=1.1\textwidth,clip]{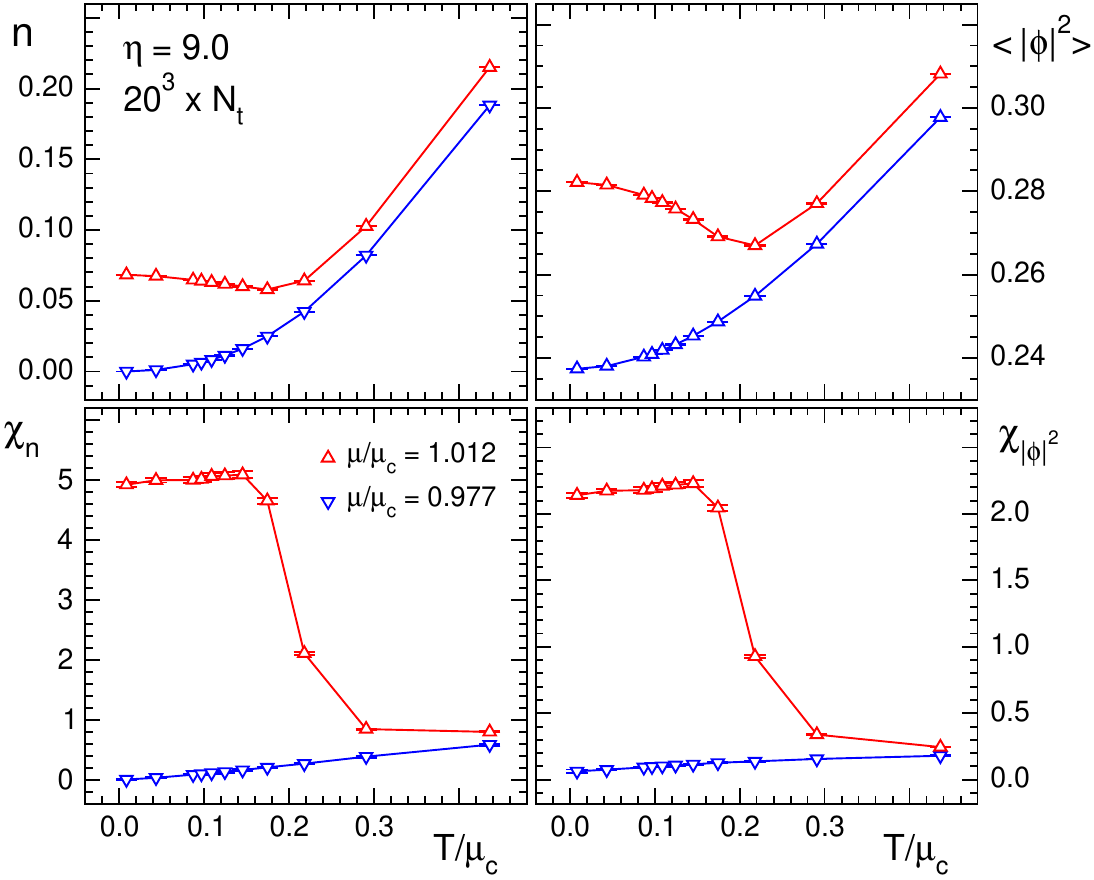}
\caption{Results for $n$, $\chi_n$, $\langle |\phi|^2 \rangle$ and $\chi_{|\phi|^2}$ for $\eta = 9.0$ and $N_s = 20$ as a function of $T/\mu_c$ for two different values of $\mu/\mu_c$. }
\label{condensation}
\end{figure}

We conclude our presentation of the finite temperature results with the discussion of observables along two vertical cuts in the phase diagram Fig.~\ref{phasediagram}. 
Using the $\eta = 9.0$ data at $N_s = 20$, in Fig.~\ref{condensation}
we show the observables $n$, $\chi_n$,  $\langle |\phi|^2 \rangle$ and $\chi_{|\phi|^2}$ as a function of $T/\mu_c$ and
compare two different values of the chemical potential. The vertical cut at $\mu/\mu_c =  1.012$ hits the phase boundary at $T/\mu_c \sim 0.2$, while the cut at $\mu/\mu_c = 0.977$ continues 
all the way to $T/\mu_c = 0$ without entering the condensed phase (compare Fig.~\ref{phasediagram}). 
This behavior is clearly reflected in the observables: For the $\mu/\mu_c = 0.977$ data the particle number density $n$ 
and the susceptibility $\chi_n$ both drop to zero as the temperature is decreased to $T/\mu_c = 0$, as expected for a value of the chemical potential in the Silver Blaze domain, i.e., 
 $\mu < \mu_c$. The vacuum expectation value $\chi_{|\phi|^2}$ drops towards 0.237, while the corresponding susceptibility $\chi_{|\phi|^2}$ vanishes as $T/\mu_c \rightarrow 0$. For the cut at $\mu/\mu_c =  1.012$ the behavior is quite different: When decreasing the temperature down to  $T/\mu_c \sim 0.2$, where the cut hits the phase boundary of the condensed phase, 
 the susceptibilities show a drastic increase an then remain essentially constant for $T/\mu_c < 0.2$. The particle number density $n$ has a non-vanishing value at $T/\mu_c = 0$ and  
$\langle |\phi|^2 \rangle$ settles at a value of  0.285. The plots nicely demonstrate the temperature driven condensation of the system for $\mu/\mu_c =  1.012$
and its absence at $\mu/\mu_c = 0.977$.

\section{Summary and discussion}
In this work we have studied a charged scalar $\phi^4$ field at finite density using the lattice formulation. The complex action problem of the standard form was 
overcome by mapping the partition sum to a flux representation where only real and non-negative terms appear. The system was updated with a generalized worm 
algorithm and we studied thermodynamical observables as a function of temperature and chemical potential.

For zero temperature the system undergoes a second order phase transition at some critical value $\mu_{c}$ that marks the end of the Silver Blaze region. 
For $\mu < \mu_{c}$ the observables are independent of $\mu$, while above a strong $\mu$-dependence sets in. The critical value $\mu_{c}$  
could very cleanly be identified from the narrow peaks of two 3rd derivatives of the free energy and in a small spectroscopy calculation we showed that $\mu_c$ 
coincides with the mass of the scalar field. A finite volume analysis 
established the second order nature of the transition. 

Increasing the temperature we found that the critical line bends towards larger values of $\mu$. For temperatures up to $T / \mu_{c} \sim 0.5$ 
we could reliably establish the continuation of second order behavior for finite $T$. At small but non-zero temperatures we found only a very weak dependence of 
observables on the chemical potential for values of $\mu$ below the critical value (pseudo Silver Blaze behavior). At the critical line we saw an abrupt change of 
physical observables marking the transition into the condensed phase characterized by strong $\mu$-dependence. When studying our observables as a function of 
the temperature we could nicely 
illustrate the temperature driven condensation for $\mu/\mu_c = 1.012$ and the absence of condensation at $\mu/\mu_c = 0.977$. 

For the analysis of temperatures  $T / \mu_{c} > 0.5$ we had to rely on finer lattices ($\eta = 7.44$) which, however, show considerably more severe finite size effects. 
For small temperatures the results match the results from our coarser ($\eta = 9.0$) lattices very well, but for $T / \mu_{c} > 0.5$ some of the observables
-- the ones related to the particle number density -- were seriously affected by finite volume effects and thus
gave inconclusive results about the nature of the transition between pseudo Silver Blaze behavior and the condensed phase with strong 
$\mu$-dependence. The observables related to derivatives of the free energy with respect to the mass parameter, i.e., 
$\langle |\phi|^2 \rangle, \chi_{|\phi|^2}$ and 
$\chi_{|\phi|^2}^{\, \prime}$, on the other hand indicate that a second order line could persist to temperatures considerably higher than  $T / \mu_{c} \sim 0.5$.
Lattice studies on large and fine lattices will be necessary to settle this issue.

\vskip7mm
\noindent
{\bf Acknowledgments:}
\vskip2mm
\noindent
We thank Gert Aarts, Shailesh Chandrasekharan, Ydalia Delgado Mercado, Hans Gert Evertz, David Kaplan, 
Anyi Li and Dam Thanh Son for discussions. C. Gattringer thanks the INT at the University of 
Washington in Seattle, where part of this work was done, for their hospitality and the Dr. Heinrich J\"org Foundation of the Karl-Franzens-University Graz for
financial support. This work was partly supported by the DFG SFB TRR 55, "Hadron Physics from Lattice QCD".

\newpage
\section*{Appendix: Pseudo-code for the worm algorithm}

In this appendix we provide pseudo-code for the initialization of the worm and its subsequent steps. The worm algorithm 
that updates the variables $k_{x,\nu} \in \mathds{Z}$ is set up in a fixed background of the variables $l_{x,\nu}$ generated with local 
Metropolis sweeps (see Section \ref{sec_worm}). The starting point of a worm will be denoted by $x_0$, the position of its head by $x$, 
and $\Lambda$ will be used for the set of all lattice sites. $\Delta$ is an equally distributed random variable with $\Delta \in \{-1,+1\}$ 
that determines whether the worm tries to increase or decrease the flux along its path. By $\nu \in \{ \pm 1, \pm 2, \pm 3, \pm 4 \}$ we 
denote the direction of the next attempted step of the worm, and in case $\nu$ is negative ($\nu = - |\nu|$) we use the convention 
$k_{x,-|\nu|} = k_{x-\widehat{|\nu|},|\nu|}$ for the assignment of the corresponding link variables. As in the main text  we use the abbreviation 
$f_x = \sum_\nu  [ |k_{x,\nu}| +  |k_{x-\widehat{\nu},\nu}| + 2( l_{x,\nu} + l_{x-\widehat{\nu},\nu})  ]$, and by $\widetilde{f}_x$ we denote the same expression but evaluated with the variable $\widetilde{k}_{x,\nu}$ from the trial offer.
Finally, {\tt rand()} denotes a random generator for uniformly distributed random numbers in the interval $[0,1)$. 
\vskip4mm
\noindent
\underline{\bf Starting the worm}
\vskip2mm
\noindent
{\tt while worm has not started} 
\vskip2mm
\noindent
\hspace*{10mm} {\tt choose} $\;\;\;\Delta \in  \{-1,+1\}\;\;\;$ {\tt randomly} \\
\hspace*{10mm} {\tt choose} $\;\;\;x_0 \in  \Lambda\;\;\;$ {\tt randomly} \\
\hspace*{10mm} {\tt choose} $\;\;\;\nu \in \{ \pm 1, \pm 2, \pm 3, \pm 4 \}\;\;\;$ {\tt randomly} \\
\hspace*{10mm} $\widetilde{k}_{x_0,\nu} = k_{x_0,\nu} + \Delta*sign(\nu) $ \\ \\
\hspace*{10mm} {\tt if} $\;\;\;|\widetilde{k}_{x_0,\nu}| \, > \, |k_{x_0,\nu}|\;\;\;$ {\tt then} \\
\hspace*{20mm} $\rho = \exp\big( sign(\nu) \, \mu \, \Delta \, \delta_{|\nu|,4}\big) \times
\big( |\widetilde{k}_{x_0,\nu}| + 
l_{x_0,\nu} \big)^{-1} \times  W(f_{x_0 + \widehat{\nu}})^{-1}$ \\
\hspace*{10mm} {\tt else} \\
\hspace*{20mm} $\rho = \exp\big( sign(\nu) \, \mu \, \Delta \, \delta_{|\nu|,4}\big) \times
\big( |k_{x_0,\nu}| + 
l_{x_0,\nu} \big) \times W(f_{x_0 + \widehat{\nu}}) ^{-1}$ \\
\hspace*{10mm} {\tt endif} \\ \\
\hspace*{10mm} {\tt if \quad rand()} $ < \rho\;\;\;$ {\tt then} \\
\hspace*{20mm} {\tt store} $f_{x_0}$ \\
\hspace*{20mm} ${k}_{x_0,\nu} \longleftarrow \widetilde{k}_{x_0,\nu} $ \\
\hspace*{20mm} $x = x_0 + \widehat{\nu}$ \\
\hspace*{20mm} {\tt worm has started} \\
\hspace*{10mm} {\tt endif} \\ \\
{\tt end while}

\vskip10mm
\noindent
\underline{\bf Running the worm}
\vskip4mm
\noindent
{\tt while} $\;\;\; x \neq x_0$  \\ \\
\hspace*{10mm} {\tt choose} $\;\;\; \nu \in \{ \pm 1, \pm 2, \pm 3, \pm 4 \}\;\;\;$ {\tt randomly} \\
\hspace*{10mm} $\widetilde{x} = x + \widehat{\nu}$ \\
\hspace*{10mm} $\widetilde{k}_{x,\nu} = k_{x,\nu} + \Delta*sign(\nu)$ \\ \\
\hspace*{10mm} {\tt if} $\;\;\;|\widetilde{k}_{x,\nu}| \, > \, |k_{x,\nu}|\;\;\;$ {\tt then} \\
\hspace*{20mm} $\rho = \exp\big( sign(\nu) \, \mu \, \Delta \, \delta_{|\nu|,4}\big) \times
\big( |\widetilde{k}_{x,\nu}| + 
l_{x,\nu} \big)^{-1} \times  W(\widetilde{f}_x) \times W(f_{x + \widehat{\nu}})^{-1}$ \\
\hspace*{10mm} {\tt else} \\
\hspace*{20mm} $\rho = \exp\big( sign(\nu) \, \mu \, \Delta \, \delta_{|\nu|,4}\big) \times
\big( |{k}_{x,\nu}| + 
l_{x,\nu} \big) \times W(\widetilde{f}_x)  \times W(f_{x + \widehat{\nu}}) ^{-1}$ \\
\hspace*{10mm} {\tt endif} \\ \\
\hspace*{10mm} {\tt if}  $\;\;\;\widetilde{x} = x_0 \;\;\;$  {\tt then} $\;\;\;\rho \longleftarrow \rho \times W(\widetilde{f}_{x_0})$ \\ \\
\hspace*{10mm} {\tt if \quad rand()} $ < \rho\;\;\;$ {\tt then} \\
\hspace*{20mm} ${k}_{x,\nu} \longleftarrow \widetilde{k}_{x,\nu} $ \\
\hspace*{20mm} $x \longleftarrow \widetilde{x}$ \\
\hspace*{10mm} {\tt endif} \\ \\
{\tt end while}

\end{document}